\documentclass[conference,10pt]{IEEEtran}
\IEEEoverridecommandlockouts
\usepackage{cite}
\usepackage{amsmath,amssymb,amsfonts}
\usepackage{soul}
\usepackage{algorithm}
\usepackage{algpseudocode}
\usepackage{graphicx}
\usepackage{textcomp}
\usepackage[font=small]{caption}
\usepackage{subcaption}
\usepackage{comment}
\usepackage{bm}
\usepackage{mathtools}
\usepackage[colorlinks=true, linkcolor=blue, citecolor=black, urlcolor=magenta]{hyperref}
\usepackage{tikz}
\usepackage{pgfplots}
\pgfplotsset{compat=1.18}

\def\BibTeX{{\rm B\kern-.05em{\sc i\kern-.025em b}\kern-.08em
    T\kern-.1667em\lower.7ex\hbox{E}\kern-.125emX}}
\usepackage{titlesec}

\titlespacing*{\section}{0pt}{5pt}{3pt}
\titlespacing*{\subsection}{0pt}{4pt}{2pt}

\newcommand{\calCN}{{\cal CN}}

\newcommand{\calK}{{\cal K}}

\newcommand{\calM}{{\cal M}}

\newcommand{\calS}{{\cal S}}
\newcommand{\calU}{{\cal U}}

\newcommand{\ds}{\displaystyle}

\newcommand{\ba}{\mathbf{a}}

\newcommand{\bd}{\mathbf{d}}
\newcommand{\bg}{\mathbf{g}}
\newcommand{\bh}{\mathbf{h}}
\newcommand{\bp}{\mathbf{p}}

\newcommand{\br}{\mathbf{r}}
\newcommand{\bs}{\mathbf{s}}
\newcommand{\bw}{\mathbf{w}}

\newcommand{\by}{\mathbf{y}}
\newcommand{\bz}{\mathbf{z}}

\newcommand{\bzero}{\mathbf{0}}

\newcommand{\bA}{\mathbf{A}}

\newcommand{\bD}{\mathbf{D}}

\newcommand{\bG}{\mathbf{G}}
\newcommand{\bH}{\mathbf{H}}
\newcommand{\bI}{\mathbf{I}}

\newcommand{\bQ}{\mathbf{Q}}
\newcommand{\bR}{\mathbf{R}}

\newcommand{\bU}{\mathbf{U}}
\newcommand{\bV}{\mathbf{V}}

\newcommand{\bbE}{\mathbb{E}}
\newcommand{\bbC}{\mathbb{C}}

\newcommand{\test}{{\underset{H_0}{\overset{H_1}\gtrless}}}
\newcommand{\norm}[1]{\left\lVert#1\right\rVert}

\begin{document}
\onecolumn
\thispagestyle{empty}

\vspace*{1cm}

\noindent
{\LARGE \textbf{IEEE Copyright Notice}}

\vspace{0.8cm}

\noindent
{\normalsize
\textcopyright\ 2026 IEEE. Personal use of this material is permitted.
Permission from IEEE must be obtained for all other uses, in any current
or future media, including reprinting/republishing this material for
advertising or promotional purposes, creating new collective works,
for resale or redistribution to servers or lists, or reuse of any
copyrighted component of this work in other works.
}

\vfill

\begin{center}
{\small
This work has been accepted for publication in the
\textit{2026 IEEE 27th International Workshop on Signal Processing
and Artificial Intelligence for Wireless Communications (SPAWC)}.
}
\end{center}

\newpage

\twocolumn
\title{Joint Detection and Velocity Estimation in OFDM-ISAC Cell-Free Massive MIMO Networks
    \thanks{The work of M. Darabi and S. Buzzi was supported by the European Commission through
the Horizon Europe MSCA Doctoral Network project  ISLANDS (grant agreement no. 101120544). The work of Sergi Liesegang was supported by  the European Commission through
the Horizon Europe MSCA Postdoctoral Fellowships project DIRACFEC (Grant No. 101108043).}
}

\author{\IEEEauthorblockN{Maryam~Darabi$^{1,2}$, Sergi~Liesegang$^{2}$, Emanuele~Grossi$^{1,2}$, and Stefano~Buzzi$^{1,2}$}
\IEEEauthorblockA{$^1$\textit{Consorzio Nazionale Interuniversitario per le Telecomunicazioni, 43124 Parma (PR) -- Italia} \\
$^2$\textit{DIEI, Universita degli Studi di Cassino e del Lazio Meridionale,
03043 Cassino, Italy} }
E-mails: \{maryam.darabi, sergi.liesegang, e.grossi, buzzi\}@unicas.it}

\maketitle

\begin{abstract}
This paper develops a sensing framework for cell-free massive MIMO (CF-mMIMO) networks operating under orthogonal frequency division multiplexing (OFDM)-based integrated sensing and communication (ISAC). The framework explicitly incorporates the 3D bistatic Doppler geometry across distributed access points (APs) into a generalized likelihood ratio test (GLRT) detector. To address system scalability, a user-target-centric AP association approach is utilized. The 3D velocity vector of the target is estimated, and several search and optimization strategies, including coarse grid search, gradient-based refinement, and particle swarm optimization (PSO), are developed and evaluated. Simulation results demonstrate that the proposed PSO‑aided detector achieves the most favorable accuracy–complexity trade‑off, while Doppler mismatch can cause substantial sensing signal‑to‑noise ratio (SNR) degradation in high‑mobility scenarios. Additionally, leveraging more OFDM subcarriers enhances frequency‑domain diversity and yields further sensing‑SNR gains.
\end{abstract}


\section{Introduction} \label{sec:1}
Integrated sensing and communication (ISAC) is a key 6G paradigm with growing relevance in high-mobility applications, such as vehicular networks \cite{liu2022integrated}, and has attracted strong interest from both academia and industry.
However, it demands careful coordination between both functions. In terms of waveform design, the orthogonal frequency division multiplexing (OFDM) waveform is widely used in ISAC thanks to its robustness to frequency‑selective fading, high‑rate support, and favorable range–Doppler coupling \cite{wang2025cooperative}. In terms of network deployment in ISAC, relying on single base station (BS) architectures inherently limit sensing degrees of freedom (DoF), degrading spatial resolution and constraining communication coverage and quality of service (QoS).  A promising solution that addresses these limitations is cell‑free massive MIMO (CF‑mMIMO) architecture \cite{buzzi2026user}. In a CF-mMIMO system, many distributed access points (APs), each with relatively simple hardware, coordinated through central processing units (CPUs), jointly serve each user equipment (UE).~This architecture provides stable signal‑to‑noise ratio (SNR), improved interference control, and coherent‑gain benefits \cite{demir2021foundations}. One of the most prominent deployment strategies for CF‑mMIMO is the user‑centric (UC) architecture \cite{buzzi2019user},  which enables scalable communication by allowing each UE to be served by a subset of APs. This concept naturally extends to the sensing functionality of an ISAC system, where only a selected subset of APs is assigned to serve a designated target \cite{liesegang2025scalable}. Moreover, a CF‑mMIMO network can emulate a bistatic or multistatic sensing configuration, thereby eliminating the need for full‑duplex operation at each AP \cite{liao2024powerallocation}.

Most ISAC studies focus on monostatic sensing configurations \cite{xu2023bandwidth,zhao2025joint,zhang2025target} or single‑BS architectures \cite{li2025efficient,xiao2024novel,zhang2025target}, which may restrict the available DoF. Aforementioned works typically estimate the radial velocity, which leaves the inherent line-of-sight (LOS) ambiguity. Although \cite{wang2025cooperative} adopts a CF‑based framework, the final implementation relies on a small number of APs, contradicting the core principles of the CF. In \cite{singh2025target}, the main emphasis is on target detection probability, and Doppler/velocity estimation is not explicitly treated. Hence, this work develops a Doppler-based detection framework under an OFDM-based ISAC-enabled CF-mMIMO architecture to address accurate Doppler estimation in vehicular scenarios and characterize the impact of Doppler mismatch on sensing performance, which, to the best of our knowledge, have~not been considered in such system configuration. We adopt a user–target-centric (UTC) approach and introduce explicit~3D bistatic Doppler profile, unlike prior studies,~enabling full 3D velocity reconstruction and reliable trajectory prediction.

The remainder of the paper is structured as follows. Section~\ref{sec:2} introduces the system model. Section~\ref{sec:3} details the sensing performance metrics. Section~\ref{sec:4} presents the numerical results, and Section~\ref{sec:5} summarizes the main conclusions.

\section{System Model} \label{sec:2}
\begin{figure}[h!]
    \centerline{\includegraphics[scale = 0.25]{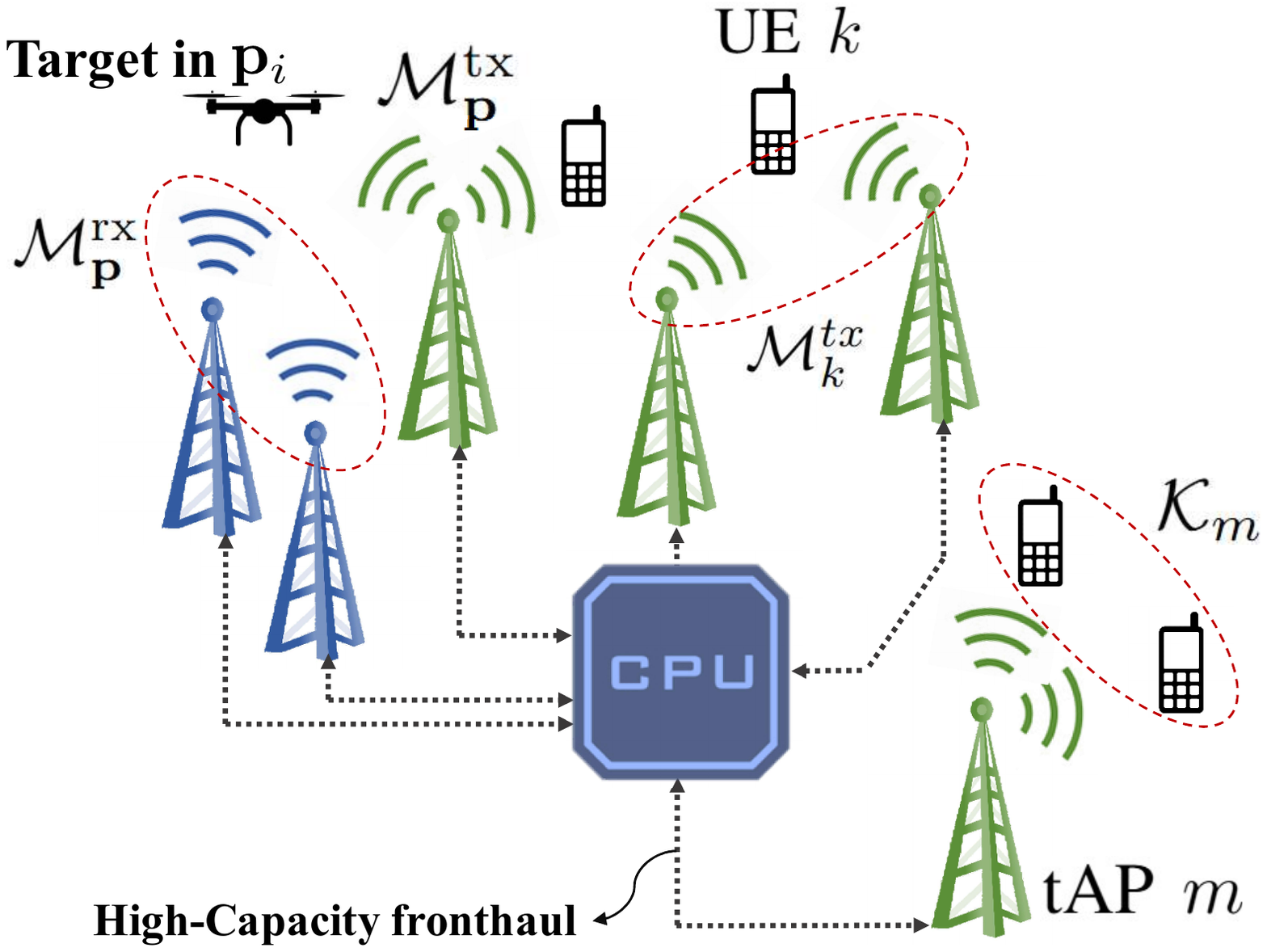}}
    \caption{ISAC-enabled CF-mMIMO network with UTC configuration.}
    \label{fig:1}
\end{figure}
We consider an ISAC-enabled CF-mMIMO network comprising $M$ distributed APs serving $K$ single-antenna UEs, along with a moving target within the coverage area. Each AP is equipped with $N_a$ antennas, with the specific array configuration left undetermined until the simulation~stage. In keeping with \cite{liesegang2025scalable}, all APs are connected to the same CPU via high-capacity fronthaul, which coordinates the distributed network. 
Fig.~\ref{fig:1} illustrates the topology of the network, where the APs are partitioned into two groups-transmit (tAPs) and receive (rAPs)- to realize a multistatic sensing architecture. Under time‑division duplexing (TDD), uplink (UL) and downlink (DL) transmissions share the same frequency band but are separated in time \cite{buzzi2017cell}. The tAPs then participate in UL training, UL/DL data transmission, and sensing beamforming, whereas rAPs exclusively handle the reception of target echoes and UL data detection. We denote the sets of tAPs and rAPs by \mbox{$\cal{M}^{\rm tx}$} and \mbox{$\cal{M}^{\rm rx}$}, respectively, with \mbox{$\cal{M}^{\rm tx} \cup \cal{M}^{\rm rx} = \cal{M}$} and {\mbox{$|{\cal{M}}|=M$}}. Following the rationale in \cite{liesegang2025scalable}, we adopt a UTC association strategy to ensure scalability in both communication and sensing (C\&S) functions. The specific UE-AP and target-AP associations are not fixed a priori for generality.

As shown in Fig.~\ref{fig:1}, a subset of tAPs, \mbox{$\calM_k^{\rm tx} \subset \calM^{\rm tx}$}, serves the $k$-th UE. Conversely, $\calK_m$ represents the set of UEs served by tAP $m$. In the sensing view, a subset of APs denoted by $\mathcal{M}_{\bp}$ inspects a given radar cell with spatial coordinates $\bp$. We define $\calM_{\bp}^{\rm tx} \triangleq \calM_{\bp}\cap \calM^{\rm tx}$ and $\calM_{\bp}^{\rm rx} \triangleq \calM_{\bp}\cap \calM^{\rm rx}$ such that $\calM_{\bp} = \calM_{\bp}^{\rm tx} \cup \calM_{\bp}^{\rm rx}$. The union of all $\calM_{\bp}^{\rm rx}$ recovers the full set of rAPs $\calM^{\rm rx}$.
Hence, a tAP may simultaneously support C\&S. The sensing task is carried out over $S$ predefined disjoint regions covering the surveillance area. Each region contains several radar range cells, and the system inspects one cell from every region at each processing interval, enabling simultaneous inspection of $S$ cells. Let \mbox{$\calS_m \subset \calS$} denote the subset of regions assigned to tAP $m$, with \mbox{$\vert \calS \vert = S$} \cite{liesegang2025scalable}.
The sensing objective is to determine the presence of a target located at \mbox{$\bp_i$}~for \mbox{$i \in S$}. 

The system operates over a total bandwidth $B$, which is jointly used for C\&S. To enable a unified and yet efficient waveform for both tasks, an OFDM modulation framework is adopted. In the frequency domain, the signal occupies $N_c$ orthogonal subcarriers with subcarrier spacing (SCS) $\Delta f$, yielding the aggregate bandwidth $B=N_c \Delta f$. From the time‑domain perspective, each OFDM symbol has duration \mbox{$T_s = T + T_{\rm CP}$}, including cyclic prefix, with the useful symbol interval $T = \frac{1}{\Delta f}$, and a total of $N_s$ OFDM symbols are transmitted within each processing block. We denote the coherence time by $\tau_c$, during which the Doppler‑induced channel variations remain approximately invariant, and the OFDM symbol satisfies \mbox{$T_s \ll \tau_c$}. This condition is consistent with requiring the maximum induced Doppler frequency to remain smaller than the SCS, i.e., \mbox{$f^d_{\rm max} \leq \Delta f$} \cite{chu2023integrated}, which ensures that phase variations across subcarriers remain negligible within one OFDM symbol.

\subsection{3D Doppler}
\begin{figure}[t]
    \centerline{\includegraphics[scale = 0.05]{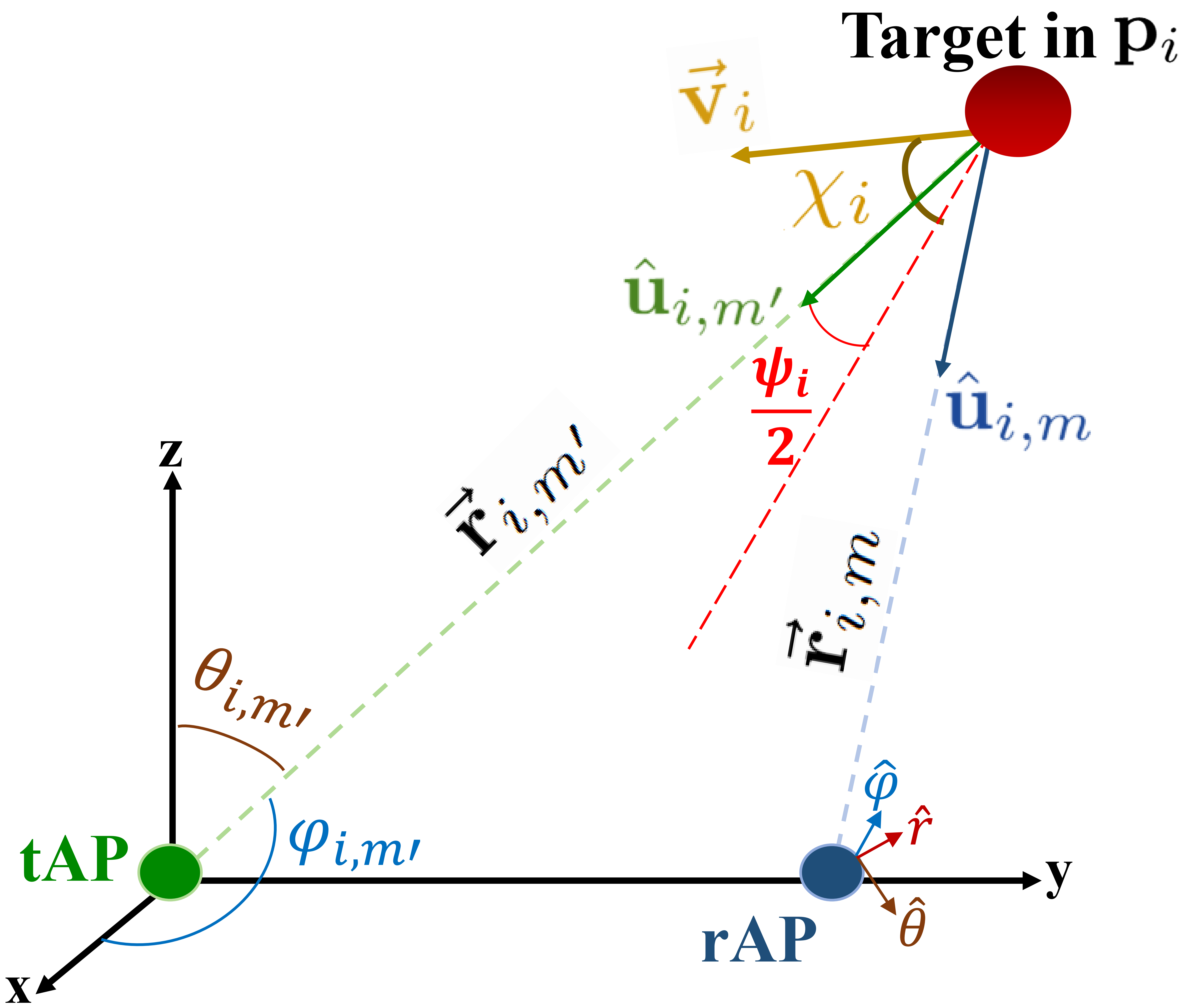}}
    \caption{Geometry for 3D-bistatic Doppler profile.}
    \label{fig:2}
\end{figure}

We extend the conventional bistatic Doppler model \cite{willis2005bistatic} to a 3D formulation incorporating the angle of departure (AOD) at the tAP and the angle of arrival (AOA) at the rAP. Let $\bp_i(t)\in\mathbb{R}^3$ denote the target position and $\vec{\mathbf v}_i(t)=d\bp_i(t)/dt$ its velocity in time $t$. As depicted in Fig.~\ref{fig:2}, let $\vec{\br}_{i,m}={\rm r}_{i,m}[\sin\theta_{i,m}\cos\varphi_{i,m}, \sin\theta_{i,m}\sin\varphi_{i,m}, \cos\theta_{i,m}]^{\rm T}$ and $\hat{\mathbf u}_{i,m}(t)$ denote the target--AP distance and corresponding unit vectors, respectively, where $\theta_{i,m}$ and $\varphi_{i,m}$ denote the elevation and azimuth AOA/AOD. The bistatic Doppler shift associated with the path $m'\!\rightarrow\! i\!\rightarrow\! m$ is then 
\begin{equation}
\begin{aligned}
     f_{i,m,m'}^d(t)&=\frac{1}{\lambda}
    \frac{d\!\left(
    \vec{\br}_{i,m'}^{\,T}(t)\hat{\mathbf u}_{i,m'}(t)
    +\vec{\br}_{i,m}^{\,T}(t)\hat{\mathbf u}_{i,m}(t)
    \right)}{dt} \\  &= \frac{1}{\lambda} \vec{\mathbf v}_{i}^{\rm T}(t)[\hat{\mathbf u}_{i,m'}(t)
    +\hat{\mathbf u}_{i,m}(t)],
\end{aligned}
    \label{eq:2D-Doppler}
\end{equation}
where $\lambda=c/f_c$, with $c$ and $f_c$ denoting the speed of light and carrier frequency, respectively. The instantaneous 3D bistatic Doppler shift can be written as
\begin{equation}
    f_{i,m,m'}^d(t)=\frac{2{\rm v}_if_c}{c}
    \cos\!\left(\frac{\psi_i(t)}{2}\right)
    \cos\!\left(\chi_i(t)\right),
    \label{eq:3Ddoppler}
\end{equation}
where ${\rm v}_i=\|\vec{\mathbf v}_i\|$. The angles $\psi_i(t)$ and $\chi_i(t)$ are given by 
\begin{align*}
\psi_i(t)
&=
\arccos\!\left(
\frac{\vec{\br}_{i,m'}^{\,T}(t)\vec{\br}_{i,m}(t)}
{\|\vec{\br}_{i,m'}(t)\|\,\|\vec{\br}_{i,m}(t)\|}
\right),\\
\chi_i(t)
&=
\arccos\!\left(
\frac{\vec{\mathbf v}_{i}^{\,T}\hat{\mathbf b}_i(t)}{{\rm v}_{i}}
\right),
\qquad
\hat{\mathbf b}_i(t)
=
\frac{\hat{\mathbf u}_{i,m'}(t)+\hat{\mathbf u}_{i,m}(t)}
{\|\hat{\mathbf u}_{i,m'}(t)+\hat{\mathbf u}_{i,m}(t)\|}.
\end{align*}

\subsection{Propagation Channels}
In the presence of a target located at $\bp_{i}$, the sensing link between the tAP $m'$ and rAP $m$ is modeled through its dominant LOS contribution. The resulting bistatic channel on subcarrier $n$ and  OFDM symbol $n'$ is 
\begin{equation}
\resizebox{0.91\linewidth}{!}{$
\bH_{i,m,m'}(n,n')= 
\tilde{\alpha}_{i,m,m'} \bA_{i,m,m'} \rho_i(n) \xi_{i,m,m'}(n') $}
\label{eq:sensing_Channel}
\end{equation} where \mbox{$n \in \{0, \ldots, N_c - 1\}$}, \mbox{$ n' \in \{0, \ldots, N_s - 1\}$}, and \mbox{$\bA_{i,m,m'}=\ba_m\left(\varphi_{m,\bp_i},\theta_{m,\bp_i}\right) \ba_{m'}^{\rm H}\left(\varphi_{m',\bp_i},\theta_{m',\bp_i}\right)$},~with \mbox{$\ba_m(\varphi_{m,m'},\theta_{m,m'})$} the array response for the azimuth \mbox{$\varphi_{m,m'}$ ($\varphi_{m',m}$)} and elevation \mbox{$\theta_{m,m'}$ ($\theta_{m',m}$)}~AOA~(AOD) for $m$-$m'$ link. The complex gain $\tilde{\alpha}_{i,m,m'}=\alpha_{i,m,m'} \sqrt{\beta_{i,m,m'}}$ encapsulates both the target's radar cross section (RCS) and the round‑trip large‑scale fading (LSF). In \eqref{eq:sensing_Channel}, $\rho_i(n) = e^{-j2\pi n \tau_i/T}$ captures the delay‑induced phase rotation and $\xi_{i,m,m'}(n')=e^{j2\pi n' f_{i,m,m'}^d T_s} $ represents the Doppler‑induced phase shift associated with the target at $\bp_i$. 
Following the Swerling‑I model, $\alpha_{i,m,m'}$ remains constant across all OFDM symbols within a coherence block \cite{behdad2024multi}, ensuring temporal stationarity of the target response over \mbox{$T_s N_s \ll \tau_c$}. 

We consider utilizing Rician fading model with random phase shift \cite{chu2023integrated} for both tAP-rAP and UE-AP links. Accordingly, the direct tAP($m'$)-rAP($m$) channel will be defined as {$\bG_{m,m'}(n) = \varkappa_{m,m'}\left(\bar{\bG}_{m,m'}(n) + \ell_{m,m'}(n)\bV_{m,m'} \right)$},~where \mbox{$\varkappa_{m,m'} = \sqrt{\beta_{m,m'}/(1 + K_{m,m'})}$} includes LSF coefficient and Rician K-factor, \mbox{$\bar{\bG}_{m,m'} \in \bbC^{N_a\times N_a}$} collects the correlated NLOS components, such that $\bar{\bg}_{m,m'} = {\rm vec}(\bar{\bG}_{m,m'}) \sim \calCN(\bzero_{N_a^2},\bQ_{m,m'})$, $\ell_{m,m'}(n) = \sqrt{K_{m,m'}}e^{j\psi_{m,m'}(n)}$ with the phase offset \mbox{$\psi_{m,m'}\sim \calU[0,2\pi]$}, and $\bV_{m,m'} \in \bbC^{N_a \times N_a}$ is the equivalent array response at the LOS direction. 

Similarly, the $N_a$-dimensional UL communication channel from UE $k$ to AP $m$ defined as {$\bh_{k,m}(n) =\sqrt{\frac{\beta_{k,m}}{K_{k,m} + 1}} \left( \sqrt{K_{k,m}} e^{j \psi_{k,m}(n)} \ba_{k,m} + \bh_{k,m}^{\text{SC}}(n) \right)$} follows the usual LOS–plus–scattering decomposition, where  $\bh_{k,m}^{\text{SC}}(n) \sim \mathcal{CN}(\mathbf{0}_{N_a}, \mathbf{Q}_{N_a})$ models diffuse fading. 

\subsection{Signal Transmission Model}
Let $x_{u}(n,n')$ denote the unit‑power OFDM symbol of stream $u$, where $u=0$ corresponds to the sensing stream and $u=1, \dots,K$ to the UEs, for the OFDM symbol index~$(n,n')$. In the DL, each tAP transmits a dual‑purpose C\&S signal. For tAP $m'$, the transmitted signal on one OFDM symbol index~$(n,n')$ is
\begin{equation}
\begin{aligned}
      \mathbf{s}_{m'}(n,n')&=\sum_{k \in {\cal{K}}_{m'}}  \sqrt{\mu_{k,{m'}}} \bw_{k,m'} x_{k,{m'}}(n,n') \\  &\quad
      + \sum_{i \in {\cal S}_{m'}} \sqrt{\eta_{i,{m'}}} \bw_{0,{m'}}(\bp_{i}) x_{0,{m'}}(n,n'),   
\end{aligned}
\label{eq:s_m}
\end{equation} here, $\mu_{k,m'}$ and $\eta_{i,m'}$ are the power allocations for C\&S, sent by tAP $m'$ to user $k$ and the potential target at position $\bp_{i}$, respectively. These coefficients satisfy the constraint $\sum_{k \in \calK_{m'}}\mu_{k,m'} + \sum_{i \in \calS_{m'}} \eta_{i,m'} \leq P_{m'}$, where $P_{m'}$ is the power budget of tAP $m'$ \cite{liesegang2025scalable}. In \eqref{eq:s_m}, $\bw_{k,m'}$ and $\bw_{0,{m'}}$ represent the precoder and beamformer for C\&S. For DL data transmission, we adopt maximum‑ratio transmission\footnote{Advanced precoding strategies (e.g., MMSE/Regularized zero-forcing) are left for future work.} \color{black} \cite{demir2021foundations}, where $\hat{\bh}_{k,m'}$ is the minimum mean square error (MMSE) channel estimate as detailed in \cite[Subsection III-C]{liesegang2025scalable}. For sensing, since the position of the inspected point is known, the beamformer is chosen as the steering vector. The two vectors are given by
\begin{align*}
    \bw_{k,m'} =\hat{\bh}_{k,m'}^*/\bigl({\mathbb{E}[\|\hat{\bh}_{k,m'}\|^2]}\bigr)^{-1/2}, \\
    \bw_{0,m'}(\bp_i) = \ba_{i,m'}(\varphi_{m',\bp_i}, \theta_{m',\bp_i}).
\end{align*}

\section{Sensing SNR} \label{sec:3}
We focus on sensing processing, since the communication side is well established in prior studies\footnote{Although the communication part will be addressed in a future extension of this work, we note that Doppler mismatch does not affect its performance.} \cite{chu2023integrated, liesegang2025scalable}.
Let $\bar{\by}_m (n,n')$ denote the $N_a$-dimensional observation collected at rAP \mbox{$m \in \calM_{\bp_i}^{\rm rx} = \calM_{\bp_i} \cap \calM^{\rm rx}$}  corresponding to the hypothesis that a target may reside at location $\bp_i$ within the $i$-th sensing region. To keep the sensing formulation focused and lightweight, we do not explicitly model clutter in this work\footnote{Clutter‑aware modeling is addressed in the extension of this work.}. That is, since all APs are synchronized through a common CPU, we assume that the AP–AP links and transmitted waveforms are perfectly known and their contributions can be canceled prior to radar processing \cite{chu2023integrated}. Thus, the residual observation has the form
\begin{equation}
   \bar{\by}_m(n,n') \approx\sum_{m' \in \calM^{\rm tx}}\bH_{i,m,m'} \bs_{m'}(n,n') +  \bar{\bz}_m(n,n'),
    \label{eq:sensing-received-signal_doppler-2}
\end{equation}
where \mbox{$\bar{\bz}_m(n,n')\sim \calCN(\bzero, \sigma^2_z \bI_{N_a})$} models thermal noise. 
Building upon the scalable hypothesis‑testing paradigm proposed in \cite{liesegang2025scalable}, our formulation extends the methodology by explicitly embedding both the per‑symbol OFDM structure and the 3D Doppler signature directly into the sensing model. Hence, we define a per‑symbol spatial–Doppler matrix as $\bD_{i,m}(n,n')=[\bd_{i,m,1}(n,n'),\dots,\bd_{i,m,|\calM^{\rm tx}|}(n,n')]$,~with $\bd_{i,m,m'}(n,n') = \sqrt{\beta_{i,m,m'}} \bA_{i,m,m'} \bs_{m'}(n,n') \rho_i(n) \xi_{i,m,m'}(n')$, and the vector of RCS coefficients is \mbox{$\bm{\alpha}_{i,m}=\left[\alpha_{i,m,1},\ldots, \alpha_{i,m,\left\vert\calM^{\rm tx}\right\vert}\right]^{\rm T}$}. 
Thus, the final observation \eqref{eq:sensing-received-signal_doppler-2} can be rewritten as below
\begin{equation}
 \bar{\by}_m(n,n') \approx \bD_{i,m}(n,n') \bm{\alpha}_{i,m} + \bar{\bz}_m(n,n'),
     \label{eq:sensing-received-signal_doppler-3}
\end{equation}
Let $\ddot{\by}_{i,m}$, $\ddot{\bD}_{i,m}$, and $\ddot{\bz}_m$ be the concatenation of observation in \eqref{eq:sensing-received-signal_doppler-3} over the entire OFDM frame $N_c \times N_s$. After collecting these samples the binary hypothesis test becomes
\begin{equation}
    \left\{\begin{array}{ll}
        H_1: & \ddot{\by}_{i,m} = \ds \ddot{\bD}_{i,m} \bm{\alpha}_{i,m} + \ddot{\bz}_m \\
        H_0: & \ddot{\by}_{i,m} = \ds \ddot{\bz}_m ,
    \end{array} \right.
    \label{eq:hypothesis}
\end{equation}
For slow target trajectories, the Doppler frequency remains approximately quasi‑static over the coherence time, yet its value is still unknown.
To address this coupling, we develop a Doppler‑based GLRT framework that jointly maximizes the likelihood over the unknown RCS vector and the Doppler frequencies. Given that $\bp_i$ is known, the delay is therefore also treated as known. This joint optimization induces a non‑convex, high‑dimensional likelihood surface, since the Doppler parameters span a 3D velocity space. Hence, no closed‑form numerical solution exists, and the computational burden scales rapidly with the granularity of the velocity grid. Building upon the formulation in \eqref{eq:hypothesis}, the corresponding generalized-likelihood-ratio-test (GLRT) takes the form  
\begin{equation}
    \frac{\max_{\bm{\alpha}_{i,m}, f_{i,m,m'}^d} f(\ddot{\by}_{i,m} | H_1, \bm{\alpha}_{i,m}, f_{i,m,m'}^d)}{f(\ddot{\by}_{i,m} | H_0)} \gtrless \delta_i.
    \label{eq:GLRT-doppler}
\end{equation} The threshold $\delta_i$ is selected to satisfy the desired probability of the false alarm for the $i$-th detection test.
The maximum-likelihood (ML) estimate of $\bm{\alpha}_{i,m}$ at a fixed Doppler frequency \( f_{i,m,m'}^d \) is $\hat{\bm{\alpha}}_{i,m} = (\ddot{\bD}_{i,m}^{\rm H}\ddot{\bD}_{i,m})^{-1} \ddot{\bD}_{i,m}^{\rm H} \ddot{\by}_{i,m}$. Substituting $\hat{\bm{\alpha}}_{i,m}$ in \eqref{eq:GLRT-doppler} and fusing all contributions at the CPU, we will have the final GLRT,
\begin{equation}
   \Lambda(\bp_i)=\max_{\vec{\mathbf v}_i} \sum\nolimits_{m \in \calM_{\bp_i}^{\rm rx}} \norm{\bU_{i,m}^{\rm H}\ddot{\by}_{i,m}}^2 \test \delta'_i,
    \label{eq:Final-GLRT}
\end{equation}with $\delta'_i=\sigma_z^{2} \ln{\delta_i}$ and $\bU_{i,m}=\ddot{\bD}_{i,m}(\ddot{\bD}_{i,m}^{\rm H}\ddot{\bD}_{i,m})^{-1/2}$. As discussed earlier, the Doppler effect is embedded in $\ddot{\bD}_{i,m}$ through $\xi_{i,m,m'}(n')$. From \eqref{eq:Final-GLRT}, the Doppler frequencies are determined by maximizing the likelihood over the 3D velocity domain defined by the bistatic geometry. To efficiently navigate this highly non‑convex Doppler‑coupled landscape, we developed a PSO‑aided, velocity‑based GLRT that jointly searches for the 3D velocity through particle updates. The PSO-aided detector operates as summarized in Algorithm~\ref{alg:PSO}. Each particle updates its velocity and position according to
\begin{equation}
    {\mathbf{u}}_p^{(\ell+1)} = w_{\ell}  {\mathbf{u}}_p^{(\ell)} + c_{1}\br_1 \odot (\bp_p - \mathbf v_p^{\ell})+ c_{2}\br_2 \odot (\bg - \mathbf v_p^{\ell}),
    \label{eq:PSO_Update1}
\end{equation}
\begin{equation}
    \mathbf v_p^{\ell+1} = \mathbf v_p^{\ell} + {\mathbf{u}}_p^{(\ell+1)},
     \label{eq:PSO_Update2}
\end{equation} where $\mathbf v_p$ is particle position (candidate target velocity), ${\mathbf{u}}_p$ is particle velocity (PSO search direction), $\bp_p$ and $\bg$ denote the personal and global bests, $w_{\ell}$ is the inertia weight, and $c_1$, $c_2$ are the so-called cognitive and social coefficients. The symbol $\odot$ denotes the Hadamard (element-wise) product. Vectors $\br_1$ and $\br_2$ contain independent and identically distributed (i.i.d.) uniform random entries. The cognitive and social terms guide each particle toward its personal and global optima. Interested readers are referred to \cite{shi1998modified} for further details. We then quantify the sensing performance at location  $\bp_i$ through a receive SNR, where the equivalence between SNR and detection probability is already found in \cite{liesegang2025scalable}. Specifically, we define the sensing SNR for the scanned cell at the $i$-th sensing region as the useful‑to‑noise power ratio in \eqref{eq:Final-GLRT} under hypothesis $H_1$, 
\begin{equation}
    \gamma_{\bp_i}= \frac{\sum_{m \in \calM_{\bp_i}^{\rm rx}} {\rm tr} \ \!\big(\bU_{i,m}^{\rm H}\ddot{\bD}_{i,m}\bR_{i,m}\ddot{\bD}_{i,m}^{\rm H} \bU_{i,m}\big)}{\!\big(\sigma_z^2 \sum_{m \in \calM_{\bp_i}^{\rm rx}} r_{i,m}\big)}
\label{eq:gamma_detection_single}
\end{equation}
where $r_{i,m}$ is the rank of $\ddot{\bD}_{i,m}$, and $\bR_{i,m}= \bbE[\bm{\alpha}_{i,m}\bm{\alpha}_{i,m}^{\rm H}]$. 
\begin{algorithm}[t]
\caption{PSO-Aided GLRT Velocity Estimation}
\label{alg:PSO}
\begin{algorithmic}[1]
\Require Cell position $\bp_i$, observations $\{\bar{\by}_m\}$.
\State Initialize $P$ particles $\{\mathbf v_p\}$ within $[\mathbf v_{\min},\mathbf v_{\max}]$.
\For{each particle}
    \State Evaluate $\Lambda(\mathbf v_p,\mathbf \bp_i)$ based on left-side \eqref{eq:Final-GLRT}.
    \State Update personal and global bests.
\EndFor
\For{$\ell=1,\ldots,L$} ($L$: number of iterations)
    \For{each particle}
        \State Update velocity and position using \eqref{eq:PSO_Update1}-\eqref{eq:PSO_Update2}
        \State Evaluate $\Lambda(\mathbf v_p,\mathbf \bp_i)$.
        \State Update personal and global bests.
    \EndFor
    \If{convergence criterion satisfied} break. \EndIf
\EndFor
\State $\hat{\mathbf v}_i\leftarrow \mathbf v_{\mathrm{gbest}}$, $\Lambda(\mathbf \bp_i)\leftarrow \Lambda(\hat{\mathbf v}_i,\mathbf \bp_i)$
\end{algorithmic}
\end{algorithm}

\section{Numerical Simulations} \label{sec:4}
We consider a 3D area scenario with a \mbox{$500 \times 500\quad \rm m^{2}$} azimuthal footprint, divided into \mbox{$S=4$} sensing regions. APs, UEs, and the target are randomly deployed; UEs and APs have fixed heights of $1.65$ m and $10$ m, respectively, while the target height is uniformly distributed between $20$ m and $100$ m. We set $M = 16$ and $K=8$. A LSF-based association policy is used, where each UE is served by $M_c=4$ APs, though more advanced policies can also be employed \cite{di2026general}. We follow the micro-urban scenario from \cite{liesegang2025scalable}, $P_{m'}=2\text{W}$, $N_0=-174 \text{dBm/Hz}$ and the receiver noise figure is 9 dB, $\alpha_{i,m,m'}\sim\mathcal{CN}(0,\sigma_\alpha^2)$ with $\sigma_\alpha^2$ set by an average RCS of $10$ dBsm. The OFDM parameters are set as $f_c =3 \text{GHz}$, $B=20\text{MHz}$, $N_c=12$ and $N_s=14$. The radar cells per region are inspected by $\calM_{\bp}^{\rm tx}=\calM_{\bp}^{\rm rx}=8$ APs. We consider a uniform power allocation scheme between UEs and target. In line with CF literature, we adopt a uniform linear array (ULA) configuration with $N_a=4$ antennas per AP. The maximum target velocity is fixed at $150$ m/s, unless otherwise stated.  

\subsection{Case Study 1} \label{subsec:A}
\begin{figure}[t]
    \centerline{\includegraphics[scale = 0.25]{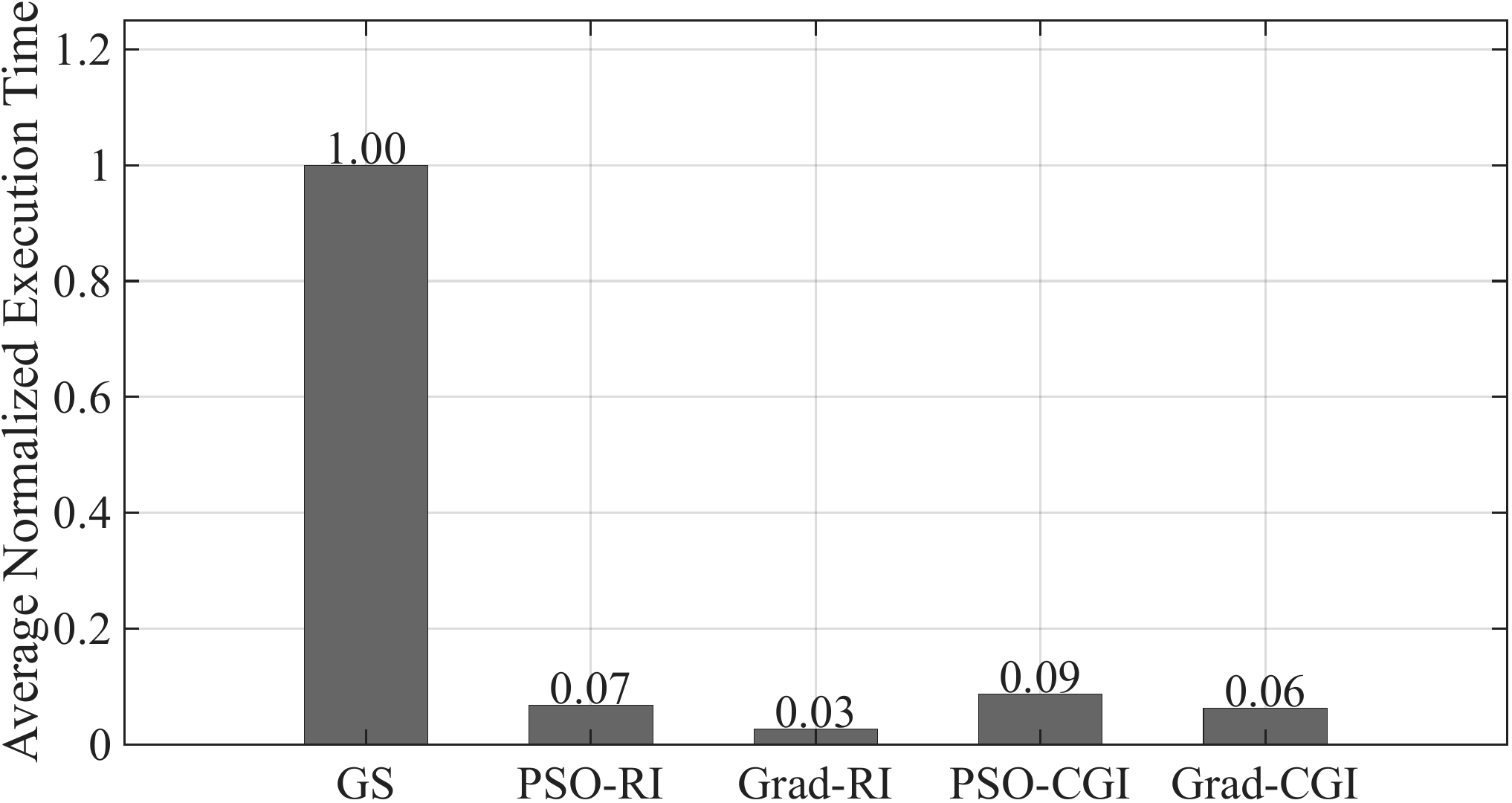}}
    \caption{Normalized average execution time of each method.}
    \label{fig:3}
\end{figure}

\begin{figure}[t]
    \centerline{\includegraphics[scale = 0.23]{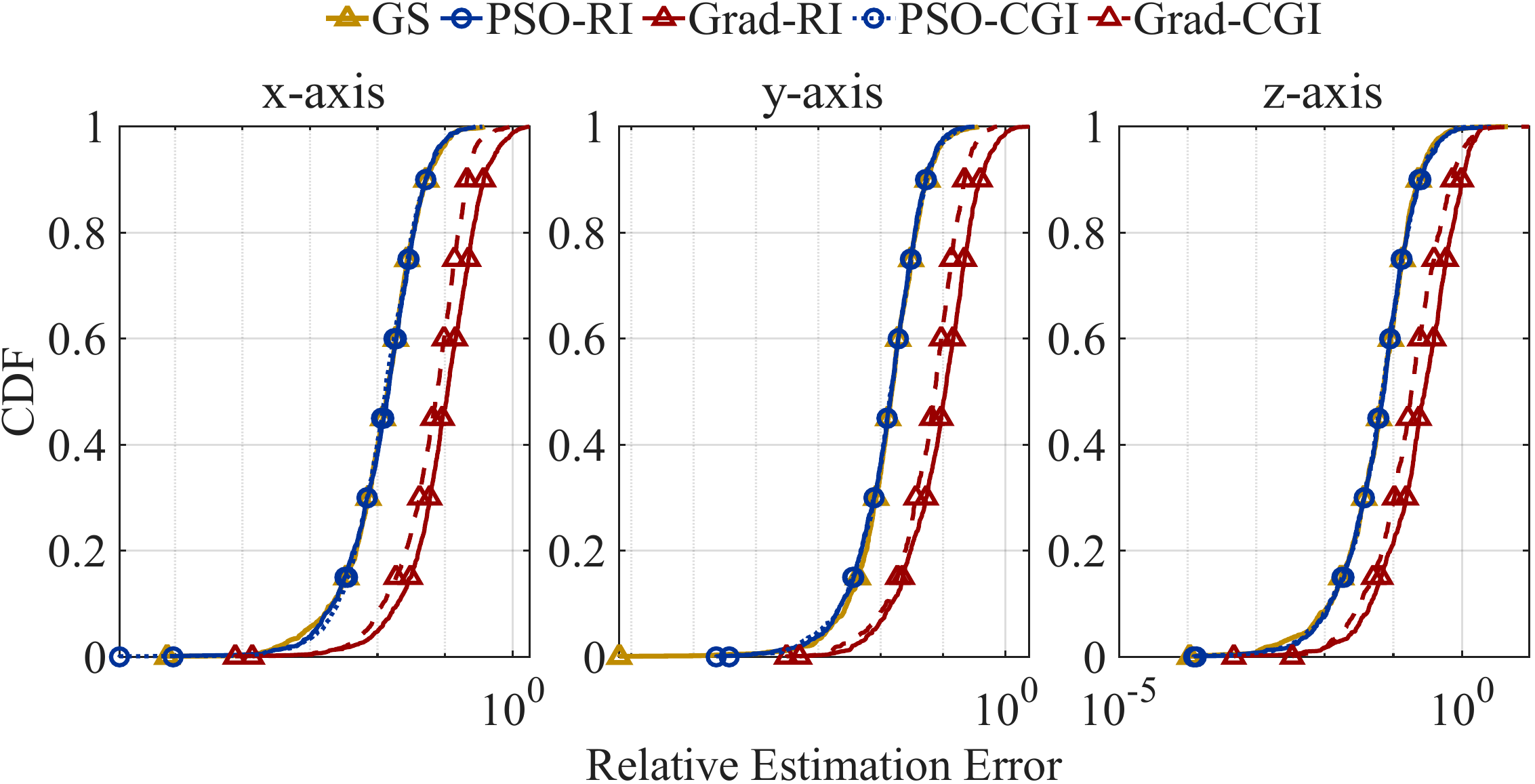}}
    \caption{CDF of the relative velocity‑estimation error across methods.}
    \label{fig:4}
\end{figure}

The distributed CF‑mMIMO architecture largely resolves velocity‑vector ambiguity, enabling full 3D velocity estimation instead of the radial‑only measurements. We conduct an experiment under 5 velocity‑estimation strategies, including PSO and gradient-ascent (Grad) methods with random and coarse-grid initializations (RI and CGI). As shown in Fig.~\ref{fig:3}, the brute‑force grid search (GS) serves as the worst‑case computational baseline, whereas PSO‑RI offers the best accuracy–complexity trade‑off. Fig.~\ref{fig:4} reports the CDF of the relative velocity‑estimation error: in $90\%$ of cases, PSO achieves ${<}5\%$ error along $\rm x$/$\rm y$ and ${<}20\%$ along $\rm z$, demonstrating strong robustness to initialization. Although CGI can accelerate PSO convergence, its coarse grid step increases overall runtime, making RI—with particles drawn uniformly over $\pm150$\,m/s per component—the more efficient choice. The reduced accuracy in the elevation component is expected, given the limited vertical sensitivity of the fixed AP height and ULA geometry.

\subsection{Case Study 2} \label{subsec:B}
\begin{figure}[t]
    \centerline{\includegraphics[scale = 0.25]{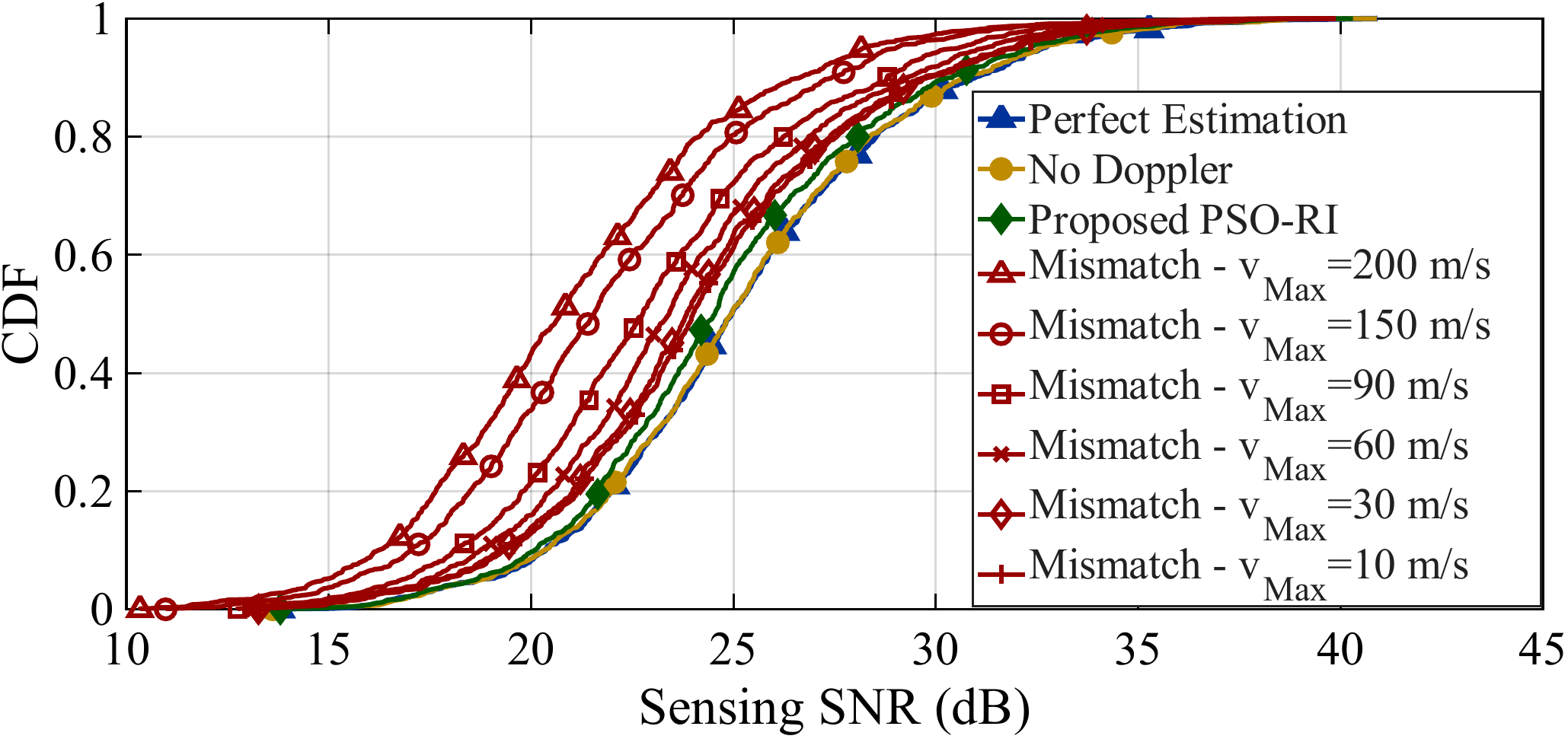}}
    \caption{Comparison of sensing SNR CDF for different Doppler cases.}
    \label{fig:5}
\end{figure}

We examine the impact of Doppler compensation by comparing 3 scenarios: (i) a stationary target with zero Doppler; (ii) a moving target whose 3D velocity $\vec{\mathbf{v}}_i$ is estimated using PSO‑RI; and (iii) the same moving target, but with the detector forced to assume a zero‑velocity vector. Fig.~\ref{fig:5} reports the resulting sensing‑SNR CDFs. The results show that accurate Doppler estimation is essential in high‑mobility ISAC settings. The proposed PSO‑RI detector preserves sensing SNR across all mobility levels, whereas neglecting Doppler—by imposing a zero‑velocity assumption—incurs an SNR degradation exceeding $5$ dB. The severity of this mismatch grows with the target’s velocity range. This performance gap highlights the decisive role of the proposed PSO‑based strategy and its suitability for high‑mobility ISAC scenarios.

\begin{figure}[t]
    \centerline{\includegraphics[scale = 0.25]{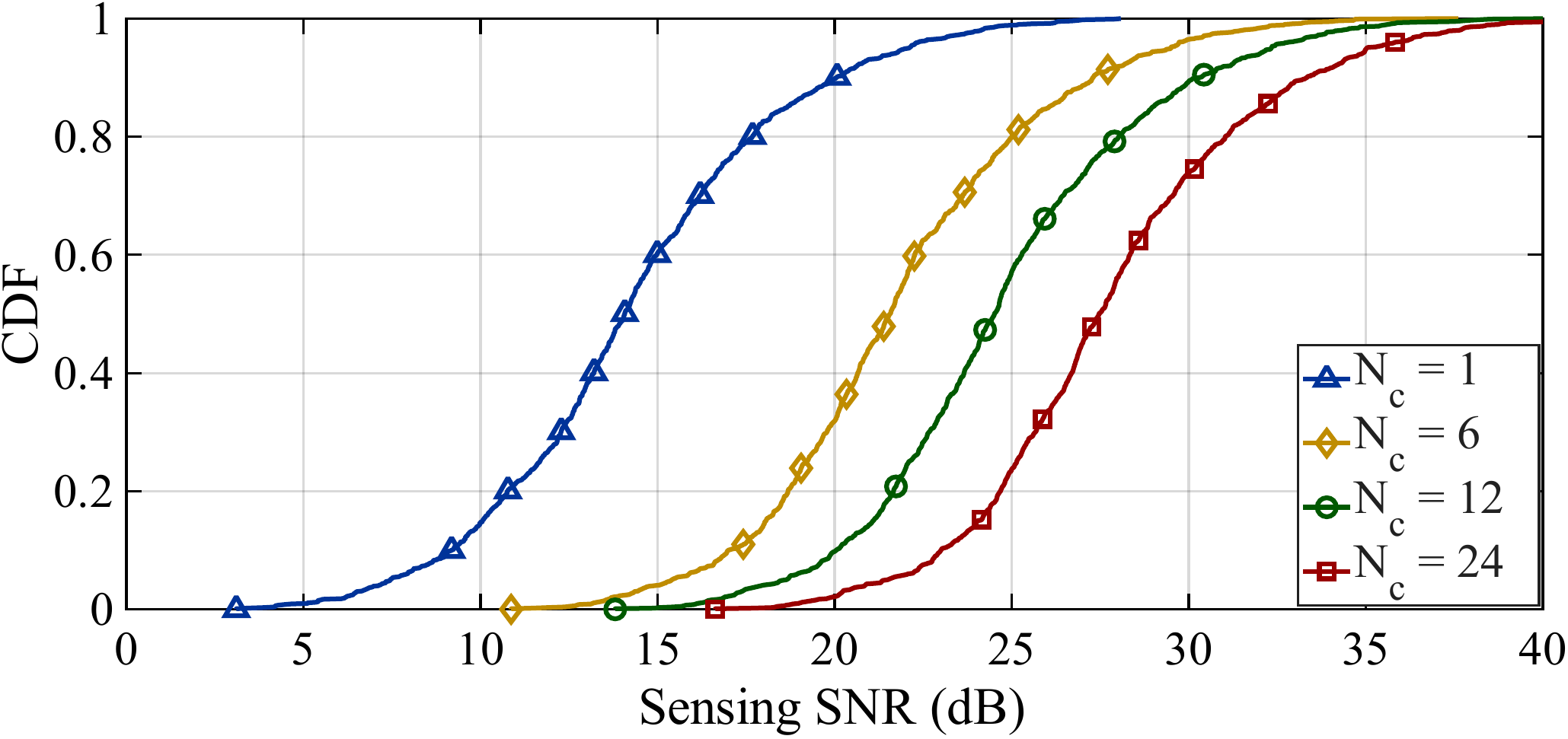}}
    \caption{CDF of sensing SNR for the PSO-aided detector versus $N_c$.}
    \label{fig:6}
\end{figure}

\subsection{Case Study 3} \label{subsec:C}
To evaluate the influence of the OFDM structure on the proposed PSO-RI detector, we analyze how the number of subcarriers affects sensing performance. Unlike many existing ISAC studies that restrict processing to a single coherence block, we evaluate different values of $N_c$, thereby exploiting time--frequency samples beyond the coherence bandwidth. With the total bandwidth and transmit power fixed, increasing $N_c$ improves the sensing SNR by providing additional frequency-domain observations rather than higher radiated power. As shown in Fig.~\ref{fig:6}, increasing from a single subcarrier to \mbox{$N_c=24$} yields more than a $10$ dB sensing SNR gain.

\section{Conclusions} \label{sec:5}
This paper analyzed the impact of Doppler mismatch in OFDM‑based ISAC CF‑mMIMO systems, showing that inaccurate or ignored velocity estimation can significantly degrade sensing performance.~The proposed PSO‑aided, Doppler‑based GLRT demonstrated strong robustness across diverse mobility conditions, reliably recovering 3D velocity even under poor initialization. Results further showed that leveraging richer OFDM resources improves sensing reliability. Overall, the framework provides a scalable and robust solution for high‑mobility ISAC scenarios.

\bibliographystyle{IEEEtran}
\bibliography{references}

\end{document}